\documentclass[a4paper,10pt]{article}
\usepackage{graphicx,amssymb,amstext,amsmath}
\usepackage{color}

\definecolor{cbl}{rgb}{0,0,1}                % bleu
\newcommand{\cbl}[1]{{#1}} 

\newcommand{\bc}{\begin{center}}
\newcommand{\ec}{\end{center}}
\def\ba#1{\begin{array}{#1}\displaystyle}
\newcommand{\ea}{\end{array}}

\newcommand{\beq}{\begin{equation}}
\newcommand{\eeq}{\end{equation}}
\newcommand{\beqa}{\begin{eqnarray}}
\newcommand{\eeqa}{\end{eqnarray}}
\newcommand{\no}{\nonumber}
\newcommand{\n}{\nonumber\\}
\newcommand{\bi}{\begin{itemize}}
\newcommand{\ei}{\end{itemize}}
\def\lt#1{\left#1}
\def\rt#1{\right#1}
\def\t#1{\tilde{#1}}

\def\frc#1#2{\frac{#1}{#2}}

\newcommand{\p}{\partial}

\newcommand{\bra}{\langle}
\newcommand{\ket}{\rangle}

\newcommand{\R}{{\mathbb{R}}}

\newcommand{\Or}{{\cal O}}

\newcommand{\ep}{\epsilon}

\newcommand{\Tr}{{\rm Tr}}

\newcommand{\rhoness}{\rho_{\text{ness}}}

\begin{document}

%\begin{titlepage}
%~ \vspace{-2.5cm}

\begin{center}

{\Large {\bf Time-reversal symmetry and fluctuation relations in non-equilibrium quantum steady states}}

\vspace{0.8cm} 

{\large \text{Denis Bernard${}^{\clubsuit}$
and Benjamin Doyon${}^{\spadesuit}$}}

\vspace{0.2cm}
{\small ${}^{\clubsuit}$ Laboratoire de Physique Th\'eorique de l'ENS, CNRS $\&$ Ecole Normale Sup\'erieure de Paris, France.}\\
{\small ${}^{\spadesuit}$ Department of Mathematics, King's College London, London, United Kingdom.}
\end{center}

\vspace{0.5cm} 
	In this note, we present a simple derivation, from time-reversal symmetry, of fluctuation relations for steady-state large deviation functions in non-equilibrium quantum systems. We further show that a condition of pure transmission implies extended fluctuation relations, connecting large deviation functions to mean currents at shifted temperatures and chemical potentials. We illustrate these concepts in various examples, including the interacting resonant level model and conformal or integrable models.

\medskip

%{\ }\hfill June 2013
\medskip

\section{Introduction}

The study of out of equilibrium phenomena is a research line of high current interest, both theoretically and experimentally \cite{review}. Of particular interest are situations where steady currents of local quantities exist: steady flows of energy, charge, particles, etc. In classical non-equilibrium systems, powerful methods adapted to studying fluctuations of these flows and their large deviation functions have led to a good understanding \cite{revue}, at least of simple model systems \cite{Jona,Derrida}. In parallel, symmetry properties of the large deviation functions, known as the fluctuation relations (of Cohen and Gallavotti) \cite{Galla} and their relatives \cite{Jarz,Croox}, have been uncovered. These relations generalize the fluctuation-dissipation relations of equilibrium systems. A further step in the understanding of out of equilibrium phenomena has been reached when realizing \cite{Maes} that the fluctuation relations are intimately related to the behaviour of the system under time reversal.

Recent experimental advances on mesoscopic systems and cold atom gases  triggered the need to better understand these concepts in the quantum realm, see e.g. \cite{ExpFluc}. Simple model systems have been theoretically analyzed where the nature of the non-equilibrium quantum steady state can be comprehended, and the large deviation function of energy and charge flows can be studied and in some cases exactly computed \cite{LL,rlmfluctu,SaitoDhar,saleur2,BD1}. \cbl{Large deviations have also been analysed in quantum quenches  \cite{GamSil13}.} In all known cases, fluctuation relations have been shown to be satisfied, and this is expected to be generic \cite{Espo,Kurchan,JarzWoj}.

The aim of this note, extending the results of \cite{JarzWoj} and of \cite{BD1,BD2}, is to present a simple derivation of fluctuation relations for non-equilibrium quantum steady states from time-reversal symmetry, and a condition of pure transmission which implies an extended version of these relations. \cbl{The validity of fluctuation relations in non-equilibrium quantum steady states is a nontrivial problem, at the basis of many recent investigations of specific models or family of models (see e.g. \cite{SaitoDhar,saleur2,BD2}).} Our setup consists in considering two quantum systems prepared at different temperatures and chemical potentials, which are put into contact at some initial time, so that energy and charge flow from one system to the other. For instance, we may have in mind two electronic leads independently equilibrated at different voltages, connected to each other on a surface through which heat and charge currents flow due to electron transfer. We further assume a quantum measurement protocol giving the definition of the large deviation function first written in \cite{LL}; it was obtained there using indirect measurements, but it is expected (see \cite{rlmfluctu,BD3,BD2}) to also follow in the steady-state limit from the two-time von Neumann measurement protocol used for instance in \cite{JarzWoj}.

Assuming that a steady state is reached at large time, we show that: \\
$\bullet$ The existence of time-reversal symmetry ensures that fluctuation relations, written in \eqref{fr}, hold for the large deviation function of energy and charge transfers in non-equilibrium steady states specified in \eqref{F}.\\
$\bullet$ A condition of pure transmission implies that the large deviation function is linked to the mean currents but at shifted temperatures and chemical potentials, see equation \eqref{efr}. This then implies that all cumulants can be obtained from the mean current alone, \cbl{and also that all cumulants, including the current, can be obtained from the {\em equilibrium} cumulants.} \cbl{The condition of pure transmission demands that energy emanating from the far left (right) goes through the system (and in particular through the contact point) towards the far right (left) without being reflected. }

These statements are nontrivial extensions of the results of \cite{JarzWoj} and  \cite{BD1,BD2} \cbl{in three ways: we deal with established non-equilibrium steady states instead of transients \cite{JarzWoj}, we use a scattering approach which technically clarifies some of the approximations done in \cite{JarzWoj}, and we point out the general consequences of the pure transmission property. The latter emerged from our recent studies \cite{BD1,BD2} on non-equilibrium conformal field theory but their domain of validity and their derivation go much beyond those particular models.} We illustrate these points in various examples involving quantum dots, the Ising chain, and conformal and integrable systems \cbl{(like the Heisenberg chain or the sine-Gordon model)}, noting in particular that in the latter three examples and with space homogeneity, heat flow is always purely transmissive  \cbl{(whence, for instance, the calculation of cumulants is much simplified)}.

\section{Framework}

\subsection{Stationary state}

Consider two quantum systems, which we will refer to as left and right systems,  with hamiltonians $H_l$ and $H_r$ and possibly two additional charges $N_l$ and $N_r$ associated to the left and right systems respectively. The systems are prepared in the Gibbs states represented by the density matrices
\[
	\rho_{l,r} := \frak{n}\lt[e^{-\beta_{l,r}(H_{l,r}-\mu_{l,r}N_{l,r})}\rt],
\]
where we use the notation $\frak{n}(A) = A / \Tr(A)$ (so that the density matrices are properly normalized). Here $\beta_{l,r}$ are the inverse temperatures of the left/right systems, and $\mu_{l,r}$ are their chemical potentials. We introduce the total hamiltonian and charge of the uncoupled systems as
\beq\label{H0}
	H_0 := H_l + H_r,\quad N_0 := N_l + N_r.
\eeq

At some initial time, the systems are connected, so that the evolution hamiltonian from that time is $H := H_l + H_r + \delta H$, for some $\delta H$ which is assumed not to commute with $H_0$ in order to generate a nontrivial dynamics. A picture that one may keep in mind is that $H_l$ and $H_r$ are extensive with energy spectrum scaling like the volume, and that $\delta H$ is not volume extensive (say it corresponds to a short-range coupling on the contact surface between the left and right systems).

The evolved density matrix of the complete system is
\[
	\rho(t) := e^{-itH} \,(\rho_l \otimes \rho_r) \,e^{itH}.
\]
We assume that after a long time a steady state is reached. Since $H_0$ commutes with the Gibbs state $\rho_l\otimes \rho_r$, the steady state density matrix is then formally
\[
	\rhoness = S_+\, (\rho_l \otimes \rho_r)\, S_+^{-1}
\]
where
\beq\label{Slim}
	S_+ := \lim_{t\to\infty} e^{-itH} \,e^{itH_0}.
\eeq
The operator $S_+$ brings states to minus infinite time with the $H_0$-dynamic and then back to time zero with the $H$-dynamics. It intertwines $H_0$ and $H$, that is $S_+H_0 = H S_+$, and we can define $N = N_l + N_r + \delta N$ such that it as well intertwines $N_0$ and $N$, that is $S_+ N_0 = N S_+$. These intertwining relations ensure that if the temperatures and chemical potentials are equal, $\beta_l=\beta_r$ and $\mu_l=\mu_r$, the stationary state is a Gibbs state for $H$ and $N$ at that temperature and chemical potential.

We expect the limit (\ref{Slim}) to exist only for infinitely large systems, and not otherwise, as in infinitely large systems the left and right parts can play the role of reservoirs. For instance, we do not expect finite systems to thermalize at large times, but rather to be subject to oscillations, and the intertwining relations cannot hold for finite systems with discrete energy spectra. Further, the limit makes sense only when taken in expectation values with local observables\footnote{In quantum field theory for instance, a local observable, by definition, has zero equal-time commutator with the energy density at positive distance.}. In order to be slightly more precise, we should rather define the steady state by its action on the algebra of operators, with an appropriate condition of locality, via the duality relation $\Tr (\rhoness \Or)=\bra \Or\ket_{\rm ness}$ with 
\[
	\bra \Or \ket_{\rm ness} = \lim_{t\to\infty} \bra e^{iHt}\, \Or\, e^{-iHt}\ket_{0}
	= \bra S_+^{-1} \,\Or \,S_+ \ket_{0}
\]
where $\bra\cdots\ket_0$ is the expectation with the density matrix $\rho_l\otimes \rho_r$. All large-time limits we will be discussing will be understood in the above sense: inside expectation values with local operators inserted. These considerations have been analyzed in details in conformal field theory in \cite{BD2}. 

Finally, we may define the operators
\[
	H^+_{l,r} := S_+ H_{l,r} S_+^{-1},\quad
	N^+_{l,r} := S_+ N_{l,r} S_+^{-1}.
\]
Note that we have $H_l^+ + H_r^+= H$ and $N_l^+ + N_r^+ = N$, thanks to (\ref{H0}) and the intertwining  relation for $S_+$. Using these operators, we can express $\rhoness$ as
\beq\label{rhoness}
	\rhoness = \frak{n}\lt[e^{-\beta_l (H_l^+-\mu_l N_l^+) -\beta_r (H_r^+-\mu_r N_r^+) }\rt].
\eeq
Contrary to $H_{l,r}$ and $N_{l,r}$, the operators $H^+_{l,r}$ and $N^+_{l,r}$ are not expected to have local properties (e.g. to have associated local densities).

\subsection{Time-reversal symmetry}

We are going to assume that there is time-reversal symmetry. More precisely, we assume the existence of an anti-linear invertible operator $\tau$, with the properties
\beq\label{tau}
	\tau H \tau^{-1} = H,\quad \tau N \tau^{-1} = N,\quad
	\tau H_{l,r} \tau^{-1} = H_{l,r},\quad
	\tau N_{l,r} \tau^{-1} = N_{l,r}.
\eeq
We  gather some information on the relation between $\tau$ and the operators involved in the construction of the steady state. Let us introduce a secondary scattering operator $S_-  = \lim_{t\to-\infty} e^{-itH} e^{itH_0}$, which, by opposition to $S_+$, transports states to plus infinite time, instead of minus infinite time, with the $H_0$-dynamics, and then backwards with the $H$-dynamics. We have $\tau \,S_\pm\, \tau^{-1} = S_\mp$. Defining also
\[
	H^-_{l,r} := S_- H_{l,r} S_-^{-1},\quad
	N^-_{l,r} := S_- N_{l,r} S_-^{-1}
\]
we see that $\tau H_{l,r}^\pm \tau ^{-1} = H_{l,r}^\mp$ and $\tau N_{l,r}^\pm \tau ^{-1} = N_{l,r}^\mp$, and, as a consequence, $H_l^- + H_r^-= H$ and $N_l^- + N_r^- = N$. This implies that, similarly to $S_+$, the operator $S_-$ intertwines $H_0$ and $H$, and also $N_0$ and $N$.

\section{Fluctuation relations from time-reversal symmetry}

\subsection{Large deviation function of energy and charge transfers}

We consider fluctuations of the energy $E$ and charge $Q$ differences,
\beq
	E := \frc12 (H_l- H_r),\quad Q := \frc12 (Q_l - Q_r).
\eeq
\cbl{Here $Q$ can be any charge as long it is associated to a local current.}
Let $R:=\lambda E + \sigma Q$. Following \cite{LL}, we define the scaled cumulant generating function (the large deviation function is by convention the Legendre transform of this, by abuse of language we will refer to $F$ as the large deviation function) by~{\footnote{\cbl{The definition of the large deviation function requires some cares. Since it is a delicate point, let us elaborate a bit on it. Quantum mechanics provides well defined rules for specifying probability distributions of measurement outputs. A quantity such as the time difference $R(t)-R$ cannot be {\it directly} measured as one cannot measure instantaneously a difference of two observables evaluated at different times. One therefore needs to specify the measurement protocol before defining the large deviation function. In \cite{LL} an indirect measurement was advocated (although maybe not in very precise way) which leads to the definition (\ref{F}) for the generating function. In \cite{Espo,Jarz} a direct two time measurement procedure was put forward, which yields another definition for the large deviation function. It is however commonly believed that the two definitions coincide once the large time limit as been taken and lead to \eqref{F}. This has been explicitly checked, for instance, in conformal field theory \cite{BD2} and in free fermion models \cite{rlmfluctu,BD3}. A drawback of the two-time definition is that it requires to projectively measure the energy or the charge on the two infinitely large subsystems at two different times, a procedure which is never realised experimentally. We would like to claim that a clearer picture would arise if one would define the large deviation function by continuously measuring the charge or energy transfer and look at the behaviours of the associated quantum trajectories. This point is clearly beyond this paper scoop, but we plan to come back to it in a near future.}}
\beq\label{F}
	F(\lambda,\sigma) := \lim_{t\to\infty} \frc1t \log
	\bra e^{i R(t)} e^{-i R}\ket_{\rm ness}
\eeq
where $R(t) = e^{itH} R\, e^{-itH}$ is the time-evolved $R$. This codes for the fluctuations of the transfer of energy and charge, $\Delta E(t) := E(t)-E$ and $\Delta Q(t) := Q(t)-Q$, at large time $t$, and their correlations \cbl{(see the previous footnote)}. Expanding $F(\lambda,\sigma)=\sum_{m,n\geq 0} C_{m,n} (i\lambda)^m (i\sigma)^n/(m!n!)$ gives $C_{m,n}$, the scaled large-time limit of the cumulants of $\Delta E(t)$ and $\Delta Q(t)$. In particular, the first power is the mean current, $J_{E} := \lim_{t\to\infty} t^{-1} \bra \Delta E(t) \ket_{\rm ness}$ and similarly for $J_Q$,
\beq
	J_E
	= -i\lt.\frc{d F}{d\lambda} \rt|_{\lambda=\sigma=0},\quad
	J_Q
	= -i\lt.\frc{d F}{d\sigma}\rt|_{\lambda=\sigma=0}.
\eeq

\subsection{Statement} \label{sectstat}

We will show that time-reversal symmetry implies the fluctuation relation for the large deviation function
\beq\label{fr}
	F(\lambda,\sigma) = F(i\gamma-\lambda,-i\omega - \sigma),
\eeq
where $\gamma:= \beta_l-\beta_r$ and $\omega := \beta_l\mu_l-\beta_r\mu_r$. \cbl{As usual, this relation implies a symmetry relation for the associated probability distribution of energy and charge transfer. We however deal with the function $F$ as the generating function for scaled cumulants and treat $i\lambda$ and $i\sigma$ as formal parameters.}

The reality of the Taylor coefficients of the function $F(\lambda,\sigma)$, expanded in powers of $i\lambda$ and $i\sigma$, is not immediate from the definition \eqref{F}, but necessary for its interpretation as a scaled cumulant generating function. Using similar tools, but without needing time-reversal symmetry, we will show that these coefficients are indeed real.

There is another ``naive'' definition of the large deviation function:
\[
	\t F(\lambda,\sigma) := \lim_{t\to\infty} \frc1t \log
	\bra e^{i (R(t)- R)}\ket_{\rm ness}.
\]
Although we cannot prove the fluctuation relations for this definition using the same tools, from \eqref{GEQ} below, it follows that both definitions agree if the following additional condition holds:
\[
	[R^+,R^-] = 0.
\]
Then $\t F(\lambda,\sigma)$ satisfies the fluctuation relations. This additional condition is equivalent to $[(S R S^{-1}),R]=0$ where $S=S_-^{-1} S_+$ is an operator that commutes with $H_0$ and which can be naturally interpreted as relating ``in'' states to ``out'' states, \cbl{respectively defined as freely propagating states in the far past and far future as usual}. We note that the particular cases $RS=\pm SR$ satisfy this additional condition. With the positive sign, this can be interpreted as pure reflection, in which case $F(\lambda,\sigma)$ is in fact independent of $\lambda$ and $\sigma$ (no flow).  The negative sign is more interesting, and can be interpreted as pure transmission. We will study this case in the next section.

\subsection{Derivation} \label{sectder}

Let us first concentrate on the simple case of energy transfer with $\sigma=\mu_l=\mu_r=0$, so that $R$ specializes to
$\lambda E = \frc\lambda2 \lt(H_l-H_r\rt)$. The stationary density matrix simplifies to
\[
	\rhoness =\frak{n}\lt[e^{-\beta_lH^+_l-\beta_rH^+_r}\rt]=\frak{n}\lt[ e^{-\beta H} e^{-\gamma E^+}\rt]
\]
where $\beta := \frc12(\beta_l+\beta_r)$ and $\gamma := \beta_l-\beta_r$, see (\ref{fr}). We use the notation $E^\pm := S_\pm E S_\pm^{-1}$. In order to split the exponential into a product of two factors, we used the fact that $[E^+,H]=0$, which holds  because $H = H_l^+ + H_r^+$, and $E$ commutes with $H_l$ and $H_r$ (we also have $[E^-,H]=0$ for similar reasons).

We first derive a formula for the $\lambda$-derivative of the large deviation functions. Let us define the energy current operator
\beq
	j_E(t) := \frc{dE(t)}{dt} = i[H,E(t)],\quad
	E(t) - E(0) = \int_0^t ds\,j_E(s).
\eeq
Using this, we find
\beqa
	-i\frc{dF(\lambda)}{d\lambda} &=&
	\lim_{t\to\infty} \frc1t \Tr\lt(
	\frak{n} \lt[e^{-i\lambda E} \,\rhoness\, e^{i\lambda E(t)}\rt] \int_0^t\,j_E(s)\,ds
	\rt) \n
	&=&
	\lim_{t\to\infty} \Tr\lt(
	\frak{n}\lt[e^{-i\lambda E(-t/2)}\, \rhoness \,e^{i\lambda E(t/2)}\rt] j_E(0)
	\rt). \label{dF0}
\eeqa
In the second line we first used time translation invariance (which follows from the stationary property of $\rhoness$ under the $H$-dynamics) in order to shift time by $-t/2$. The integral of the current is then transformed into $\int_{-t/2}^{t/2} j_E(s)\,ds$, which, in the large-$t$ limit, is dominated by its bulk part around $s=0$ of extent  $t + O(1)$. This bulk part is $s$-independent as we will see shortly, which gives rise to a factor of $t$. This cancels the factor $1/t$ and produces the expression on the second line. In arguing that the bulk part dominates and that it is of extent  $t+O(1)$, we  assumed locality for the current as discussed above, \cbl{and, also as above, the fact the observables local to the contact point reach a steady regime so that at long times all transients only contribute negligibly to the integral} (we do not elaborate more on this in order not to obscure the main logical steps). We note that such a locality property holds for instance in all examples below.

We may now take the large-$t$ limit, and at the same time justify the $s$-independence of the bulk part. If in \eqref{dF0} we would have keep  the current $j_E$ at some fixed time $s$, after time translation we would have to consider $E(-s\pm t/2)$. Since $E$ commutes with $H_0$, we immediately obtain
\beq\label{limit}
	\lim_{t\to\infty} E(-s\pm t/2) = 
	e^{-isH}\lt(\lim_{t\to\pm \infty} e^{itH/2} e^{-itH_0/2} \,E\, e^{itH_0/2} e^{-itH/2}\rt)e^{isH}
	= e^{-isH}\,E^\mp\,e^{isH} = E^\mp
\eeq
(where we also used the fact that $E^\mp$ commutes with $H$), from which it follows that
\beq\label{dF}
	-i\frc{dF(\lambda)}{d\lambda}=\Tr\lt(
	\frak{n} \lt[e^{-i\lambda E^+}\, \rhoness\, e^{i\lambda E^-}\rt] j_E(0)
	\rt).
\eeq
Recall that our system is of infinite extent, and there, the equations in (\ref{limit}) do not make sense as operatorial equations: \cbl{for an infinite system, the total energies on the left and right, and the total energy transferred at infinite time, are infinite.} However, as discussed above, the limit and the operators involved make sense when taken in a normalized density matrix with insertions of local operators, as in \eqref{dF0} and \eqref{dF}. For a discussion in the context of conformal field theory, see \cite{BD2}.

Rearranging the right-hand side of (\ref{dF}) using $[H,E^\pm]=0$, and defining $G:=-i\frc{dF}{d\lambda}$, we have
\beq\label{G1}
	G(\lambda)=-i\frc{dF(\lambda)}{d\lambda}=\Tr\lt(
	\frak{n} \lt[e^{-\beta H}\,e^{i\lambda' E^+}\, e^{i\lambda E^-}\rt] j_E(0)
	\rt)
\eeq
where $\lambda' = i\gamma-\lambda$.

We now use the time-reversal symmetry. We observe that $\tau j_E(0) \tau^{-1} = -j_E(0)$ and that $\tau E^\pm \tau^{-1} = E^\mp$. Since $\tau$ is anti-unitary, $\overline{\Tr(A)} = \Tr(\tau A \tau^{-1})$ for any linear operator $A$. This implies
\[
	\overline{G(\lambda)} =
	-\Tr\lt( 
	\frak{n} \lt[e^{-\beta H}\,e^{-i\,\overline{\lambda'} E^-}\, e^{-i\,\overline\lambda E^+}\rt] j_E(0)
	\rt).
\]
Taking the hermitian conjugation in order to get rid of the complex conjugation, using $\overline{\Tr(A)} = \Tr(A^\dag)$, and to interchange the two factors involving $E^\pm$, we get
\beq
	G(\lambda) 
	= -\Tr\lt( 
	\frak{n} \lt[e^{-\beta H}\, e^{i\lambda E^+}\,e^{i\lambda' E^-}\,\rt] j_E(0)
	\rt).\label{G2}
\eeq
Here we used the fact that $H$ commutes with $E^\pm$. Comparing (\ref{G1}) and (\ref{G2}), there is equality, up to a sign, under exchanging $\lambda\leftrightarrow\lambda'$, that is $G(\lambda) = - G(i\gamma-\lambda)$. By integration this implies
\beq\label{frenergy}
	F(\lambda) = F(i\gamma-\lambda).
\eeq

The same derivation applies in the general case with $R= \sigma Q + \lambda E$ and the stationary state given by
\[
	\rhoness = \frak{n}\lt[e^{-\beta (H - \mu N) + \omega Q^+ - \gamma E^+}\rt]
\]
where $\beta\mu := \frc12(\beta_l\mu_l + \beta_r\mu_r)$ and we recall that $\omega = \beta_l\mu_l-\beta_r\mu_r$. Introducing the charge current $j_Q:= i[H,Q]$ as well as the $\sigma$-derivative $G_Q := -i \frc{d F}{d\sigma}$, and using similar manipulations as those above, we get
\beq
	G_{E,Q}(\lambda,\sigma) = \Tr\lt(
	\frak{n} \lt[e^{-\beta (H - \mu N)}\, e^{i\lambda' E^+ + i\sigma' Q^+}\, e^{i\lambda E^- + i\sigma Q^-}\rt] j_{E,Q}(0)
	\rt) \label{GEQ}
\eeq
where $\lambda' = i\gamma-\lambda$ and $\sigma' = -i\omega-\sigma$. Time-reversal  invariance then implies
\[
	G_{E,Q}(\lambda,\sigma) = -G_{E,Q}(i\gamma-\lambda,-i\omega - \sigma)
\]
which by integration again implies the fluctuation relation \eqref{fr}.

Finally, we provide a proof of the reality of the Taylor coefficients, using some of the above arguments (but not using time-reversal symmetry). We concentrate again on the case $\sigma=\mu_l=\mu_r=0$ for simplicity, but the general case is done similarly. Consider
\[
	f(\lambda) := \lim_{t\to\infty} \frc1t \Tr\lt(\frak{n}\lt[e^{-\beta H - \gamma E}\rt]
	e^{i\lambda E(t)} \,e^{-i\lambda E}\rt).
\]
By the above arguments (taking derivative, shifting time by $-t/2$ by translation invariance and evaluating the large time limit using \eqref{limit}), we get $-i df(\lambda)/d\lambda = G(\lambda)$, whence $f(\lambda) = F(\lambda)$. But also, for $\lambda\in\R$,
\beqa
	\overline{f(-\lambda)} &=& \lim_{t\to\infty} \frc1t \Tr\lt(\frak{n}\lt[e^{-\beta H - \gamma E}\rt]
	e^{-i\lambda E} \,e^{i\lambda E(t)} \rt) \n
	&=&
	\lim_{t\to\infty} \frc1t \Tr\lt(\frak{n}\lt[e^{-\beta H - \gamma E}\rt]
	 e^{i\lambda E(t)}\,e^{-i\lambda E(i\beta)} \rt)
\eeqa
where in the second step we have used the cyclic property of the trace. 
Hence
\[
	-i \frc{d\,\overline{f(-\lambda)}}{d\lambda} =\lim_{t\to\infty} \frc1t
	\Tr\lt(\frak{n}\lt[
	e^{-i\lambda E(i\beta)} \,
	e^{-\beta H - \gamma E}\,e^{i\lambda E(t)}\rt]
	\lt(\int_{0}^{t} j_E(s)\,ds+\int_{i\beta}^{0} j_E(s)\,ds\rt)\rt).
\]
The second integral inside the parentheses gives a vanishing contribution in the limit $t\to\infty$, because it is finite, hence killed by the factor $1/t$. The  first integral is treated in the usual way: shifting by $-t/2$, and using locality and (11). To deal with the factor containing $E(i\beta)$ we may make the replacement $s\to s-i\beta$ in (11). Then we find again the right-hand side of \eqref{G1}. This shows that $\overline{f(-\lambda)} = f(\lambda)$ for $\lambda$ real, whence the reality of the Taylor coefficients in the expansion in powers of $i\lambda$.

\subsection{Example}

To illustrate the concepts above, let us consider the interacting resonant-level model (IRLM) \cite{irlm}. This describes the coupling of two fermionic leads via a two-level quantum dot. The hamiltonian of this system is $	H = H_0 + H_{\rm int}$ with
\beqa\label{H}
	H_0 &:=& -i v_F\int_{-\infty}^\infty dx(\psi_1^\dag \p_x \psi_1 + \psi_2^\dag\p_x \psi_2)
	+ \ep\, d^\dag d\n
	H_{\rm int} &:=& \eta ( (\psi^\dag_1(0)+\psi^\dag_2(0)) d + d^\dag (\psi_1(0)+\psi_2(0)))
	 + U (\psi_1^\dag(0)\psi_1(0) + \psi_2^\dag(0)\psi_2(0) ) d^\dag d \no
\eeqa
where $d^{\dag}$ is the fermion creation operator on the dot, $\{d^{\dag},d\}=1$, $\epsilon$ the dot energy, $\psi_j(x)$, $j=1,2$, are the fermionic operators of the leads, $\{\psi_j^{\dag}(x),\psi_k(y)\}=\delta_{j;k}\delta(x-y)$, $v_F$ is the Fermi velocity, $\eta$ the hopping amplitude and $U$ is the Coulomb interaction parameter. The above description is the ``unfolded'' one, where the left and right leads are described by pure right movers $\psi_1$ and $\psi_2$. Hence we identify $H_{l,r} = -i v_F\int_{-\infty}^\infty dx\,\psi_{1,2}^\dag \p_x \psi_{1,2}$.

One can check that the system is time-reversal symmetric, under the anti-unitary operator $\tau$ defined by
\[
	\tau \psi_{1,2}(x) \tau^{-1} = \psi_{1,2}(-x),\quad
	\tau d \tau^{-1} = d.
\]
This leaves $H$ and $H_{l,r}$ invariant. Our above derivation then proves the fluctuation relations for the IRLM at any values of the couplings. A strategy to compute the large deviation function using Bethe ansatz has been proposed in \cite{saleur1}, and the fluctuation relation has been checked \cite{CBS11,saleur2}. In the case $U=0$, the large deviation function for charge transfer is known and given by the Levitov-Lesovik formula \cite{LL}, which satisfies the fluctuation relation.

In the free-fermion case $U=0$, the model can be solved exactly. The $\tau$-symmetry, and its implementation in the fluctuation relations, then has a nice geometrical interpretation. We can diagonalize the system by writing\footnote{We use the notation of \cite{BD3}.}
\begin{eqnarray} \label{psi_mode}
	\psi_1(x,t) + \psi_2(x,t) &=& \int \frc{dp}{\sqrt{\pi}}\, e_p(x)e^{-ipt}\, a_p ,\nonumber\\
	\psi_1(x,t) - \psi_2(x,t)) &=& \int \frc{dp}{\sqrt{\pi}}\, e^{ip(x-t)}\, b_p,\\
	-2i\eta\, d(t) &=& \int \frc{dp}{\sqrt{\pi}}\, w_p e^{-ipt}\, a_p,\nonumber
\end{eqnarray}
with $a_p$ and $b_p$ canonical fermionic operators, $\{a_p^\dag,a_{p'}\} = \{b_p^\dag,b_{p'}\} = \delta(p-p')$ (other anti-commutators vanishing). Here $e_p(x) := e^{ipx}$ for $x>0$ and $e_p(x) := v_p e^{-ipx}$ for $x<0$, where $v_p = e^{i\phi_p}$ for some known scattering phase $\phi_p$ \cite{irlm} and $w_p := v_p-1$. With the above expression \eqref{psi_mode}, the $\tau$ operator acts on the modes as
\beq\label{tauap}
	\tau a_p \tau^{-1} = v_p \,a_p,\quad \tau b_p \tau^{-1} = b_p.
\eeq

Let us consider for now the charge fluctuations only, which have been studied in the past \cite{rlmfluctu,BD3}. We have $H = \int dp\,E_p \,(a^\dag_p a_p + b^\dag_p b_p)$ where $E_p=p$ is the energy of the modes, and $N = \int dp\,(a^\dag_p a_p + b^\dag_p b_p)$. The solution also gives explicit expressions for the stationary state and for $Q^+$ \cite{BD3}, whose $\tau$ transform gives $Q^-$, that is $Q^+ =
\frc12\int dp\,(a^\dag_p b_p + b^\dag_p a_p)$ and $Q^- =\frc12 \int dp\,(v_p^\dag \,a_p^\dag b_p + v_p \,b_p^\dag a_p)$. The charge current is given explicitly by $j_Q(0) \propto {\rm Im}(d^\dag \psi_o(0)) \propto 	{\rm Im} ( \int dpdq\, e^{-i\phi_p/2} \sin\phi_p \,a^\dag_p b_q)$.
Then, computing $G_Q(\sigma)$ as a trace on the Fermi Fock space involves only operators bilinear in the modes. They are all diagonal in the momentum space except for $j_Q(0)$, and the non-diagonal part of $j_Q$ gives zero contribution by conservation of fermion number in each $p$-space. The trace becomes a product over $p$ of traces over two-dimensional one-particle spaces  (made of particles of type $a_p$ and $b_p$) of operators whose $p$-dependent representations are, up to normalizations,
\[
	H \mapsto E_p\, \mathbb{I},\quad N \mapsto \mathbb{I},\quad
	Q^+ \mapsto \sigma_x ,\quad
	Q^- \mapsto  \sigma_{\phi_p},\quad
	\lt. j_Q(0)\rt|_{\rm diag} \mapsto \sigma_{\phi_p/2+\pi/2}.
\]
Here $\sigma_\theta$ is the Pauli matrix in the direction at angle $\theta$ with respect to the $x$ axis, $\sigma_\theta = \lt(\begin{smallmatrix}0 & e^{-i\theta} \\ e^{i\theta} & 0\end{smallmatrix}\rt)$.

We can now picture geometrically the $\tau$ transformation in these one-particle spaces. Indeed, on the above operators the $\tau$ symmetry (\ref{tauap}) acts as a reflexion with respect to the line at angle $\phi_p/2$, which corresponds to $\sigma_x \leftrightarrow \sigma_{\phi_p}$ and $ \sigma_{\phi_p/2+\pi/2} \mapsto - \sigma_{\phi_p/2+\pi/2}$. This is in agreement with  $\tau Q^\pm\tau^{-1}= Q^\mp$ and $\tau j_Q(0) \tau^{-1}= -j_Q(0)$.

We note finally that it is a simple matter to generalize the Lesovik-Levitov  formula to include both charge and energy transfers. Recall that this formula has the form $F_{LL}(\sigma) = \int dp\,F^p_{LL}(\sigma;\omega)$ for some explicitly known functions $F^p_{LL}$ \cite{LL}, which depend on the difference of chemical potentials via $\omega$. Using explicit forms for $E^\pm$ and the energy current $j_E$ similar to those for the charge, we observe that the formula is modified, in each momentum sector, by the changes $\sigma \mapsto \sigma + \lambda E_p$ and $\omega\mapsto \omega- \gamma E_p$, so that the complete large deviation function is $F(\lambda,\sigma) = \int dp\,F^p_{LL}(\lambda,\sigma;\omega)$ with
\begin{eqnarray} \label{LLgeneral}
	F^p(\lambda,\sigma;\omega) = F_{LL}^p(\sigma+\lambda E_p;\omega - \gamma E_p).
\end{eqnarray}
It is a simple matter to check that the fluctuation relations (\ref{fr}) for $F(\lambda,\sigma)$ then follows from the usual charge fluctuation relations for $F_{LL}(\sigma)$.

\cbl{Eq.(\ref{LLgeneral}) applies as well to situations where the charge quanta have different dispersion relations $E_p$ and to more species of fermions.}

\section{Pure transmission and consequences}

\subsection{Statement}

We now consider a slightly stronger condition that leads to ``extended'' fluctuation relations, whereby the derivatives of the large deviation function are the mean currents at shifted temperatures and chemical potentials. Let us concentrate again first on the energy fluctuations only. We do not need to assume time-reversal symmetry. We assume rather that there is pure transmission; that is, with $S = S_-^{-1} S_+$ the operator relating ``in'' and ``out'' states introduced in Subsection \ref{sectstat}, we require that
\beq\label{extra}
	SE = - ES.
\eeq
\cbl{The scattering operator by definition describes how the parts of the excitations situated in the far left (right) that travel towards the right (left) interact (reflection, transmission) with each other over time to produce excitations in the far left (right) traveling towards the left (right). The pure transmission condition \eqref{extra} indicates that the total energy in the far left (right) of right- (left-) moving excitations goes through the whole system towards the far right (left) without being reflected.}

Equation \eqref{extra} is equivalent to $E^+=-E^-$, i.e. $\tau E^- \tau^{-1} = - E^-$ (inverting time is compensated by exchanging the role of left and right sectors, $l\leftrightarrow r$). Although this may look like a strong requirement, there are actually many important examples where it is fulfilled (see below). From \eqref{G1}, it then immediately follows that
\beq\label{efr}
	-i\frc{d F(\lambda)}{d\lambda}
	= J_E(\beta_l+i\lambda,\beta_r-i\lambda)
\eeq
where $J_E(\beta_l,\beta_r)$ is the mean energy current at inverse temperatures $\beta_l$ and $\beta_r$. This is the {\em extended fluctuation relations} (in the case of energy transfer). In other words, the $d/d\gamma$ derivatives ($\gamma = \beta_l-\beta_r$) of the mean current itself give the higher cumulants,
\[
	F(\lambda)\;=\; \sum_{n=1}^\infty \frc{(2i\lambda)^n}{2\,n!} \frc{d^{n-1}}{d\gamma^{n-1}} J_E \;=\; \cbl{\int_0^{i\lambda} dz\, J_E(\beta_l+z,\beta_r-z).}
\]
From expression \eqref{efr}, the fluctuation relations \eqref{frenergy} follow simply if the current is anti-symmetric under the exchange of the left and right temperatures: $J_E(\beta_l,\beta_r) = -J_E(\beta_r,\beta_l)$. This anti-symmetry occurs for instance if there is parity symmetry. Hence, in the case of pure transmission (\ref{extra}), the extended fluctuation relations are satisfied, and the fluctuation relations hold if either time-reversal symmetry, or parity symmetry, is present.

\cbl{From \eqref{efr}, the large deviation function is obtained exactly from the current alone. Further, with $\beta_l=\beta_r=\beta$ and $i\lambda=z$, the current at temperatures $\t\beta_l = \beta+z$ and $\t\beta_r = \beta-z$ is obtained from the {\em equilibrium} large deviation function at temperature $\beta$. Also, \eqref{efr} suggests that the probability $P_{t;\beta_l,\beta_r}(q)$ of transfer of an energy $q$ has the asymptotic form $\propto e^{\frc{\beta_l-\beta_r}2 q} \,p_{t,\beta_l+\beta_r}(q)$, although this requires appropriate analytic properties of $F$ as a function of temperatures. Finally, \eqref{efr} gives $2G_E = \beta^2 {\cal N}_E$ relating the non-equilibrium differential conductance $G := \beta^2 d J_E / d\gamma$ to the second cumulant (noise) ${\cal N}_E$.}

This generalizes to include charge fluctuations as well, under the additional condition $SQ=-QS$, i.e. $Q^+ = -Q^-$: both derivatives of the large deviation function are then given by the corresponding mean currents but at shifted temperatures and chemical potentials: $\beta_{l,r}\to \beta_{l,r}\pm i\lambda$ and $\beta_{l,r}\mu_{l,r}\to\beta_{l,r}\mu_{l,r} \mp i\sigma$.

We note that in the case of pure reflection, for instance $SE=ES$ (i.e. $E^+=E^-$), expression (\ref{G1}) and (\ref{GEQ}) show that $F$ is independent of $\lambda$, which naturally means that the current and its fluctuations all vanish.

\subsection{Examples}

The simplest example is that of one-dimensional conformal field theory out of equilibrium \cite{BD1}, where we have two isomorphic conformal field theories on half lines independently thermalized, forming the left and right systems, coupled to each other at their endpoints making a homogeneous conformal system. In this case, the energy and momentum densities separate into right and left movers $h_\pm(x)$. These evolve in a chiral way, $h_\pm(x,t) = h_\pm(x\mp t)$, but the $H$ and $H_{l,r}$ dynamics differ by the boundary conditions \cite{BD1}. The action of $\tau$ is simply 
\[
	\tau \,h_\pm(x)\, \tau^{-1} = h_\mp(x).
\]
One can check that this action preserves $H$ and $H_{l,r}$, and their corresponding time evolution; for instance $\tau h_+(x,t) \tau^{-1} = \tau h_+(x-t) \tau^{-1} = h_-(x-t) = h_-(x,-t) = (\tau h_+(x) \tau^{-1})(-t)$, whence $\tau$ indeed corresponds to a time reversal. One can also check that both the $H$ and $H_{l,r}$ boundary conditions are preserved. Here, $E = \frc12(H_l-H_r)$ where $H_{l,r}:=\pm\int_0^{\pm\infty} dx\,(h_+(x)+h_-(x))$, and using $h_\pm(x,t) = h_\pm(x\mp t)$ along with the continuity of $h_\pm(x)$ at $x=0$ (under the $H$-dynamics), or directly the scattering operators $S_\pm$ described in \cite{BD2}, we find
\[
	E^+ = \frc12 \int dx (h_+(x) - h_-(x)) = -E^-.
\]
Hence we have the extended fluctuation relations \eqref{efr}, as was observed in \cite{BD1,BD2}.

We have a similar situation in the case of two Ising chains connected together in the same fashion, as was noticed in \cite{DeLuca}.  Also, we note that the extended fluctuation relations are satisfied in the $U=0$ IRLM, but only when $\eta\to\infty$ (where $v_p=-1$ and the transmission coefficient $|w_p/2|$ is 1).

Finally, it turns out that the transmission condition \eqref{extra} also holds in general integrable relativistic quantum field theory (IQFT). We consider again a similar setup, where independently thermalized left and right parts are connected at one point making a homogeneous system \cite{D}. Let us assume for simplicity that the spectrum of particles is composed of only one particle type. The Hilbert space can be described both by the basis of ``in'' and the basis of  ``out'' asymptotic states, with associated creation and annihilation operators $A_{\rm in/out}^\dag(\theta)$ and $A_{\rm in/out}(\theta)$. If $\Phi(x,t)$ is a bosonic ``fundamental field'' associated to the asymptotic particles, we define time-reversal symmetry by $\tau \Phi(x,t)\tau^{-1} = \Phi(x,-t)$ (for Dirac fields, left and right components of the Lorentz representation are exchanged). This preserves any local Hamiltonian $H$ on the real line, as well as independently $H_l$ (on the left, $x<0$) and $H_r$ (on the right, $x>0$). Further, by the standard asymptotic-state construction, this implies
\[
	\tau A_{\rm in/out}(\theta) \tau^{-1} = A_{\rm out/in}(-\theta). 
\]
Let $N_{\rm in/out}(\theta) := A_{\rm in/out}^\dag(\theta) A_{\rm in/out}(\theta)$ be the occupation numbers. The operators $H_{l,r}^+$ were found in \cite{D}, and applying $\tau$ this gives
\beq
	H_l^{\pm} = \int_{\gtrless 0} d\theta\,E_\theta \,N_{\rm in/out}(\theta)  ,\quad
	H_r^{\pm} =
	\int_{\lessgtr0} d\theta\,E_\theta\, N_{\rm in/out}(\theta)
\eeq
where ``in" corresponds to $H_{l,r}^+$ and ``out" to $H_{l,r}^-$. In IQFT, the scattering is elastic, hence the set of out-particle momenta is the same as the set of in-particle momenta \cite{Zamo}. That is $N_{\rm in}(\theta) = N_{\rm out}(\theta)$ on every asymptotic state (hence this holds as operators). This immediately implies $H_l^+ = H_r^-$ so that $E^- = - E^+$.
Hence for every IQFT, the energy fluctuations in this setup (including homogeneity) can be obtained from the energy current alone. Note that one expects that the latter may be evaluated using thermodynamic Bethe ansatz \cite{prepa}. Transfer of any local higher-spin conserved charge can be dealt with in the same manner.

In these three examples the extended fluctuation relations are not expected to hold if there is an impurity at the contact point (that produces both transmission and reflection). However, the usual fluctuation relations still hold, except if the impurity breaks the time-reversal symmetry.

\medskip

{\em Acknowledgments.} This work was in part supported by ANR contract ANR-2010-BLANC-0414.
D.B. thanks the department of mathematics at King's College London for hospitality.

\end{document}